\documentclass{webofc}
\usepackage[varg]{txfonts}   
\begin{document}
\title{Hyperonic equation of state at finite temperature \\ for neutron stars}
%
%

\author{\firstname{Hristijan } \lastname{Kochankovski}\inst{1,2}\fnsep\thanks{\email hriskoch@fqa.ub.edu{}} \and
        \firstname{Angels} \lastname{Ramos}\inst{1}\fnsep \and
        \firstname{Laura} \lastname{Tolos}\inst{3,4,5}\fnsep
}

\institute{Departament de F\'{\i}sica Qu\`antica i Astrof\'{\i}sica and Institut de Ci\`encies del Cosmos, Universitat de Barcelona, Mart\'i i Franqu\`es 1, 08028, Barcelona, Spain
\and
Faculty of Natural Sciences and Mathematics-Skopje, Ss. Cyril and Methodius University in Skopje, Arhimedova, 1000 Skopje, North Macedonia 
\and
Institute of Space Sciences (ICE, CSIC), Campus UAB,  Carrer de Can Magrans, 08193 Barcelona, Spain
\and
Institut d'Estudis Espacials de Catalunya (IEEC), 08034 Barcelona, Spain
\and
Frankfurt Institute for Advanced Studies, Ruth-Moufang-Str. 1, 60438 Frankfurt am Main, Germany
          }

\abstract{%
We review the composition and the equation of state of the hyperonic core of neutron stars at finite temperature within a relativistic mean-field approach. We make use of the new FSU2H$^*$ model, which is built upon the FSU2H scheme by improving on the $\Xi$ potential according to the recent analysis on the $\Xi$ atoms, and we extend it to include finite temperature corrections. The calculations are done  for a wide range of densities, temperatures and charge fractions, thus exploring the different conditions that can be found in proto-neutron stars, binary mergers remnants and supernovae explosions. The inclusion of hyperons has a strong effect on the composition and the equation of state at finite temperature, which consequently would lead to significant changes in the properties and evolution of hot neutron stars.
}
%
\maketitle
\section{Introduction}
\label{intro}
The large densities in the neutron star core make neutron stars perfect candidates for testing different models of nuclear matter under extreme conditions. Since there is no experimental data that covers the supranuclear region, many uncertainties are still present. One of them is the composition of the inner core, that is strongly model dependent. Numerous models predict the appearance of exotic particles, such as hyperons, as it is  energetically favourable to produce them inside the core. Although evolved neutron stars are cold objects, some of the information we have from them come from neutron star merger events or from their early evolution. In both scenarios, the matter in the star is hot, which means that a finite temperature treatment of matter is essential. In addition, one cannot always assume that star matter satisfies the beta equilibrium condition.  Due to the difficulty in dealing with finite temperature nuclear models, many groups that perform complicated relativistic simulations for the evolution of hot neutron stars use approximated models for the finite-temperature equation of state (EoS) \cite{Hotokezaka2013RemnantWaveform}. However, we have shown in Ref.~\cite{Kochankovski2022EquationMatter} that those approaches are particularly inaccurate when one considers hyperons in the core of the neutron star. Thus, it is of fundamental importance to make use of a baryonic EoS at high densities without further approximations when dealing with neutron star mergers and/or the early stages in the evolution of a neutron star. In this work we briefly present the FSU2H$^*$ model that was recently constructed in Ref.~\cite{Kochankovski2022EquationMatter}. Within this framework we build an hyperonic EoS that covers a wide range of densities, temperatures and charge fractions, making this EoS available for relativistic neutron star simulations. We show the composition pattern and the pressure-density relation for matter at different temperatures and charge fractions.

\section{Theoretical framework}
\label{sec-1}
We consider matter made of baryons at a given temperature $T$, baryon  density $\rho_B$ and fixed charged fraction $Y_Q = \sum_{i} q_i \rho_i/\rho_B$, where $q_i$ and $\rho_i$ are the charge and density of the $i$-th baryon. The interaction between the baryons is modeled through the exchange of different mesons, which leads to the following Lagrangian density $\cal{L}$ of the system that can be split into the baryonic contributions ${\cal L}_b$ 
($b$ =  $n, p,\Lambda$, $\Sigma$, $\Xi$), and a mesonic term ${\cal L}_m$, which includes the contributions from the $\sigma$ , $\omega$, $\rho$, $\phi$ and $\sigma^{*}$ mesons:
\begin{eqnarray}
{\cal L} &=& \sum_b {\cal L}_b + {\cal L}_m; \\
{\cal L}_b &=& \bar{\Psi}_b(i\gamma_{\mu}\partial^{\mu} -q_b{\gamma}_{\mu} A^{\mu} - m_b  
+ g_{\sigma b}\sigma + g_{\sigma^{*}b}  \sigma^{*} - g_{\omega b}\gamma_{\mu} \omega^{\mu}\nonumber 
- g_{\phi b}\gamma_{\mu} \phi^{\mu} - g_{\rho,b}\gamma_{\mu}\vec{I}_{b}\vec{\rho\,}^{\mu})\Psi_b, \nonumber \\
{\cal L}_m &=& \frac{1}{2}\partial_{\mu}\sigma \partial^{\mu}\sigma - \frac{1}{2}m^2_{\sigma}\sigma^2 - \frac{\kappa}{3!}(g_{\sigma b}\sigma)^3 - \frac{\lambda}{4!}(g_{\sigma b})^4 
 \nonumber \\ 
&+& \frac{1}{2}\partial_{\mu}\sigma^{*} \partial^{\mu}\sigma^{*}  -\frac{1}{2}m^2_{\sigma^{*}}{\sigma^{*}}^2 \nonumber 
-\frac{1}{4}\Omega^{\mu \nu}\Omega_{\mu \nu}  
+\frac{1}{2}m^2_{\omega} \omega_{\mu} {\omega}^{\mu} +  \frac{\zeta}{4!} g_{\omega b}^4 (\omega_{\mu}\omega^{\mu})^2 \nonumber \\
&-&\frac{1}{4}\vec{R}^{\mu \nu}\vec{R}_{\mu \nu} + \frac{1}{2}m^2_{\rho}\vec{\rho}_{\mu}\vec{\rho\,}^{\mu}+ 
\Lambda_{\omega}g^2_{\rho b}\vec{\rho_{\mu}}\vec{\rho\,}^{\mu} g^2_{\omega b} \omega_{\mu} \omega^{\mu} - \frac{1}{4}P^{\mu \nu}P_{\mu \nu}\nonumber
\\&+&\frac{1}{2}m^2_{\phi}\phi_{\mu}\phi^{\mu}-\frac{1}{4}F^{\mu \nu}F_{\mu \nu};
\label{eq:lagrangian}
\end{eqnarray}
with $m_i$ being the mass of $i$-th particle, $\Psi_b$ indicating the baryon Dirac fields, and $\Omega_{\mu \nu} = \partial_{\mu} \omega_{\nu} -\partial_{\nu} \omega_{\mu} $, $\vec{R}_{\mu \nu} = \partial_{\mu} \vec{\rho_{\nu}} - \partial_{\nu} \vec{\rho_{\mu}} $, $P_{\mu \nu} = \partial_{\mu} \phi_{\nu} -\partial_{\nu} \phi_{\mu} $ and $F_{\mu \nu} = \partial_{\mu} A_{\nu} -\partial_{\nu} A_{\mu}$ being the mesonic and electromagnetic strength tensors. The quantity $\vec{I}_b$ represents the isospin operator, $\gamma^{\mu}$ are the Dirac matrices and $g_{mb}$ labels the couplings of the different baryons to the mesons. 

Between the different baryonic species a weak interaction equilibrium is assumed that give rise to the following relations between their chemical potentials ($\mu_i$):
\begin{eqnarray}
&&\mu_{b^0} = \mu_n , \nonumber \\
&&\mu_{b^{-}} = 2\mu_n - \mu_p , \nonumber \\
&&\mu_{b^{+}} = \mu_p,
\end{eqnarray}
where $b^0$ is any neutral baryon, $b^{+}$ is any positive and $b^{-}$ is any negative baryon.

In order to obtain the composition and all thermodynamical variables of interest, one needs to obtain the Euler-Lagrange equations of motion and solve them in relativistic mean field approximation. Then, from the energy-momentum tensor it is easy to obtain all thermodynamical quantities. For details see Ref. \cite{Kochankovski2022EquationMatter}. The coupling constants are chosen in order for the model to reproduce the saturation properties of nuclear matter and some properties of finite nuclei, as well as to satisfy the constraints on the high-density nuclear pressure coming from heavy-ion collisions together with the 2 $M_{\odot}$ neutron star observations and radii smaller than 13 km \cite{Demorest2010ShapiroStar,Antoniadis:2013pzd,Fonseca2016,Cromartie2020RelativisticPulsar, Riley:2019yda,Miller:2019cac,Riley:2021pdl,Miller:2021qha}. We list the values of the parameters in Tables (\ref{table:1},\ref{table:2}), that are obtained in Ref.~\cite{Tolos2017TheStarsb,Kochankovski2022EquationMatter}. The hyperonic couplings to the mesonic fields are given in terms of their ratios to the corresponding couplings of nucleons: $R_{i Y}=g_{iY}/g_{iN}$ for $i={(\sigma,\omega,\rho)}$, and $R_{\sigma^* Y}=g_{\sigma^* Y}/g_{\sigma N}$ and $R_{\phi Y}=g_{\phi Y}/g_{\omega N}$, since $g_{\sigma^* N}=0$ and $g_{\phi N}=0$ due to the OZI rule.
\begin{table*}
\centering
\small
\caption{Parameters of the model FSU2H*. The mass of the nucleon is equal to $m_N = 939$ MeV.}
\begin{tabular}{cccccccccccccc}
\hline\noalign{\smallskip}
$m_{\sigma}$  & $m_{\omega}$ & $m_{\rho}$&$m_{\sigma^{*}}$ & $m_{\phi}$ &$g_{\sigma N}^2$  & $g_{\omega N}^2$ & $g_{\rho N}^2$ & $\kappa$ & $\lambda$ & $\zeta$ & $\Lambda_{\omega}$ \\
(MeV) &  (MeV)&  (MeV)&  (MeV)& (MeV) & & & & (MeV) & & &\\
\noalign{\smallskip}\hline\noalign{\smallskip}
498 & 783 & 763 & 980  & 1020 & 103 & 170 & 197 & 4.00014 & -0.0133 & 0.008 & 0.045 \\
\noalign{\smallskip}\hline
\end{tabular}
\label{table:1}

\end{table*}

\begin{table}
\small

\centering
\caption{\centering The ratios of the couplings of hyperons to mesons with respect to the nucleonic ones.}
\label{table:2}     
\begin{tabular}{cccccc}
\hline\noalign{\smallskip}
$Y$ & $R_{\sigma Y}$ &$R_{\omega Y} $ & $R_{\rho Y}$& $R_{\sigma^{*}Y}$ & $R_{\phi Y}$ \\
\noalign{\smallskip}\hline\noalign{\smallskip}
$\Lambda$ & $0.6613$ & $2/3$ & $0$ & $0.2812$ & $-\sqrt{2}/3$  \\
$\Sigma$ & $0.4673$ & $2/3$ & $1$ & $0.2812$ & $-\sqrt{2}/3$  \\
$\Xi$ & $0.3305$ & $1/3$ & $1$ & $0.5624$ & $-2\sqrt{2}/3$  \\
\hline\noalign{\smallskip}
\end{tabular}
\end{table}

\section{Composition and EoS at fixed $Y_Q$}

As mentioned before, we show the composition and pressure-density relation [$P(\rho_B)$] for  matter at fixed $T$ and $Y_Q$. Given that simulations need a wide range of values of these parameters, we display our results for two temperatures ($T = 25$ MeV and $T = 75$ MeV) and two charged fractions ($Y_Q = 0.01$ and $Y_Q = 0.5$) in order to give a clear picture about what one might expect for baryonic neutron star matter at any arbitrary condition.

\begin{figure*}[h]
\centering
\includegraphics[width=8cm,clip]{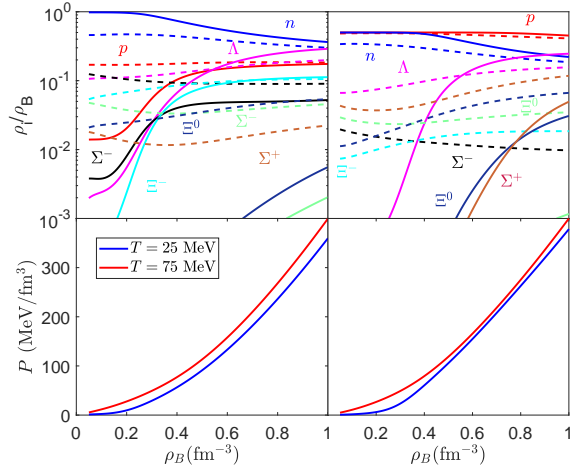}

\caption{Composition and pressure as a function of baryonic density of  the hyperonic core of neutron stars at $Y_Q = 0.01$ (left panels) and $Y_Q = 0.5$ (right panels). The solid lines in the upper plots are the compositions obtained at $T = 25$ MeV, while the dashed lines are the compositions at $T = 75$ MeV.}
\label{fig-1}       
\end{figure*}

In the upper plots of Figure \ref{fig-1} we show the composition of the core for two charge fractions ($Y_Q = 0.01$ (left panel) and $Y_Q = 0.5$ (right panel)), and two different temperatures (solid lines for $T = 25$ MeV and dashed lines for $T = 75$ MeV). We observe that when the temperature is large, hyperons are present at any point of the core. All hyperons of the baryon octet can be found in the hyperonic core for certain conditions of temperature and charge fraction. The appearance of the $\Lambda$ hyperon significantly lowers the neutron abundance, an effect which is even more enhanced with the appearance of the negatively charged hyperons. The presence of negatively charged hyperons then leads to an increase of the proton fraction due to the fixed charge fraction. This mechanism is more clear at low and moderate temperatures, such as $T = 25$ MeV. 
In the lower plots of the Figure \ref{fig-1} we display the pressure versus the baryonic density of the core for the same charge fractions and temperatures. One clearly sees that the relative difference between the pressure-density curves for different temperatures is larger for low densities in the core. This is due to the fact that temperature effects are manifest more clearly when matter is non-degenerate. However, it is interesting to note that when the charge fraction is high, as in $Y_Q = 0.5$, the difference between the curves for different temperatures becomes small at densities around $\rho_B = 0.5$ fm$^{-3}$. The reason lies in the fact that, due to the fixed charged fraction, the proton abundance cannot be significantly lowered, making the protons degenerate, even at high temperature.

\section{Conclusions}
We have briefly reviewed the EoS of hot dense baryonic matter at different charge fractions within the new FSU2H$^*$ model. All hyperons from the baryon octet can appear in the core of neutron stars in significant abundances under certain conditions of temperature and charge fraction. At sufficiently high temperature, hyperons populate the dense matter even at densities below the saturation one. This composition pattern affects the EoS, making it softer with respect to the pure nucleonic one. The correct treatment of the hyperonic EoS at finite temperature is of special importance for the correct description of the early phases in the evolution of neutron stars or for the numerical simulations of binary neutron stars mergers.
\bibliography{references.bib}
%
%
%
%

\end{document}